\documentclass[aps,prd,groupedaddress,showpacs]{revtex4}

\usepackage[parfill]{parskip}    % Activate to begin paragraphs with an empty line rather than an indent
\usepackage{graphicx}
\usepackage{amssymb}
\usepackage{epstopdf}
\usepackage{color}
\usepackage{float}
\usepackage{amsmath}

\usepackage{graphicx}
\usepackage{pdfpages}
\usepackage[colorlinks=true,citecolor=blue,urlcolor=magenta,breaklinks]{hyperref}
\begin{document}

\title{Charged Throats in the Ho\v{r}ava--Lifshitz Theory}

\author{Alvaro Restuccia}
\email{alvaro.restuccia@uantof.cl}
\affiliation{Departamento de F\'isica, Facultad de Ciencias Básicas, Universidad de Antofagasta, Casilla 170, Antofagasta, Chile.}

\author{Francisco Tello-Ortiz}
\email{francisco.tello@ua.cl}
\affiliation{Departamento de F\'isica, Facultad de Ciencias Básicas, Universidad de Antofagasta, Casilla 170, Antofagasta, Chile.}

%\date{}
\begin{abstract}
A spherically symmetric solution of the field equations of the Ho\v{r}ava--Lifshitz  gravity--gauge vector interaction theory is obtained and  analyzed. It describes a charged throat. The solution exists provided a restriction on the relation between the
mass and charge is satisfied. The restriction reduces to the Reissner--Nordstr\"{o}m one in the limit in which the coupling constants tend to the relativistic values of General Relativity. We introduce the correct charts to describe the solution across the entire manifold, including the throat connecting an asymptotic Minkowski space--time with a singular 3+1 dimensional manifold. The solution external to the throat on the asymptotically flat side tends to the Reissner--Nordstr\"{o}m space--time at the limit when the coupling parameter, associated with the term in the low energy Hamiltonian that manifestly breaks the relativistic symmetry, tends to zero. Also, when the electric charge is taken to be zero the solution becomes the
spherically symmetric and static solution of the Ho\v{r}ava--Lifshitz gravity.     
\end{abstract}


\maketitle

%%%%%%%%%%%%%%%%%%%%%%%%%%%%%%%%%%%%%%%%%%%%%%%%%%%%%

\section{Introduction}

The Ho\v{r}ava--Lifshitz (HL hereinafter) theory stands nowadays as a strong candidate to describe a quantum description of the gravitational interaction \cite{r1,r4}. The proposal breaks the relativistic symmetry by introducing an anisotropic scale of temporal and spatial coordinates. In addition, it introduces terms with higher order spatial derivatives in the potential, compatible with the geometric structure of the formulation, which make the theory, in the UV regime, renormalizable by power counting. Although the original proposal given in \cite{r1} was incomplete, since it suffered from  instabilities at all energy scales and strong coupling \cite{r4,r5,r6}, a subsequent analysis \cite{r7} overcame 
these problems by including a new terms in the Hamiltonian compatible with the symmetries of the original proposal. The anisotropic behavior becomes manifest by the introduction of the so--called critical dynamical exponent $z$. Furthermore, the exponent $z$ describes different scale energies, for example the case $z=1$ corresponds to the IR point whilst $z=3$ the UV one \footnote{Obviously, the value of the exponent $z$ at the UV point depends on the spatial dimension $d$.}. The HL theory can be seen as general framework containing two classes of theories: i) the projectable ones, namely when the lapse function is restricted to be a purely time function $N=N(t)$ and ii) the non--projectable ones where the lapse function is more general $N=N(t,\vec{x})$. Both theories are not equivalent since they do not propagate the same spectrum, besides its constraints structure is also different. 
Along the years the feasibility of the HL theory has been widely studied, for example in \cite{r8,r9,r15} the Hamiltonian formulation was analyzed, and from the phenomenological point of view in \cite{r14} the radiation of gravitational waves of self--gravitating binary systems in the low--energy limit was investigated.
On the cosmological background within the framework of the projectable version, interesting solutions have been obtained \cite{r23,r24,r25,r26} and also in the standard model domain \cite{r27} and four--fermion Gross--Neveu like models \cite{r28} novels results were presented.  

In a more particular context than the quantum and cosmological aspects that this proposal offers, it is especially interesting as well as intriguing to investigate the possible existence of exotic structures such as wormholes (WH from now on). Since the pioneering work by Morris and Thorne \cite{m1,m2}, the study of WH in the GR framework has laid the foundations for both the geometric shape that they must adopt, and the type of matter that must support them.  WHs are visualized as connecting two relatively distant regions or universes located on an asymptotically flat space--times. This type of WH is known in the literature as traversable or passable WH \cite{m1,m2,m3}. Of course, this is conditioned to a strong restriction, which implies the violation of the energy conditions \cite{m3}, mainly the null (NEC): $\rho+p_{r}\geq 0$ and $\rho+p_{t}\geq 0$, and weak (WEC): $\rho\geq0$, $\rho+p_{r}\geq 0$ and $\rho+p_{t}\geq 0$ conditions, being $\rho$ the energy--density, $p_{r}$ and $p_{t}$ the radial and lateral pressures, respectively. This type of matter is called as \emph{exotic matter} \cite{m1,m2,m3}. In the  references \cite{m4,m5} it was shown that in the background of GR it is not possible to obtain a passable WH satisfying at the same time the energy conditions. In this respect, during last years dozens of articles were devoted in obtaining WH solutions within the arena of modified gravity theories satisfying the energy conditions. For example, in the background of $f(R)$ gravity cosmological WH solutions and WH space--times satisfying different equations of state \cite{n1,n2,n3,n4,n5,n8,n6,n7} (and references therein) were studied. However, as occurs in GR, in the $f(R)$ gravity arena it is not possible to get a traversable WH space--time satisfying the energy conditions, unless the theory propagates ghost fields \cite{n9,n10}. On the other hand, in the $f(R,T)$ gravity \cite{n11,n12,n13,n14,n15}, Brans--Dicke gravity \cite{n16,n17}, Einstein--Cartan theory \cite{n18} and extra dimensions \cite{n19,n20} scenarios, it is possible to meet at the same time both the traversability and energy conditions. This is so, because the extra terms or fields introduced in these kind of theories can be seen as corrections to the Einstein field equations.

In considering the HL theory, these objects have been widely explored by taking into account both the projectable and non--projectable versions. For instance, in the projectable case, in \cite{wh1} traversable WHs connecting two asymptotically de Sitter/anti--de Sitter regions were studied, in \cite{wh2} asymptotically flat WHs threading by phantom energy were obtained. More recently, the so--called Euclidean WHs have been analyzed in the Euclidean path integral approach to quantum gravity \cite{wh3}, also within the branch of 2d causal dynamical triangulations \cite{wh4} this kind of object were explored. On the other hand, in the non--projectable branch it has been shown that these configurations
are not allowed, instead of it one obtains a throat \cite{r31,r32}. The main difference between a WH and a throat, relies on the fact that the former is connecting necessarily two asymptotically flat regions. In the models found in \cite{r31,r32} the structure is asymptotically flat only in one side, whilst is divergent at the other one. Interestingly, the throat solution does not need to be supported by any kind of exotic matter distribution, that is, this solution was obtained by solving only the fields equations without any assumption on the energy--momentum tensor \cite{r31} (vacuum solution). 

Taking advantage of these remarks, in the present paper we study charged throats within the non--projectable HL background. The motivation behind this analysis, is supported on recent articles \cite{r29,r30} working out the anisotropic gravity--vector gauge coupling in 3+1 dimensions. This formulation is obtained, starting from the  the 4+1 dimensional non--projectable HL theory after a dimensional reduction \emph{a la} Kaluza--Klein. The resulting 3+1 dimensional theory exhibits the same features as the original proposal \cite{r1} \i.e., the anisotropic gravity-vector gauge coupling respects the $\text{Diff}_{\mathcal{F}}(\mathcal{M})$ symmetries \cite{r29,r30} and is power--counting renormalizable \cite{arxiv}. In this case, in contrast with the solution provided in \cite{r31}, here we have a spherical throat supported by and static spatial electric field. This contribution is coming from the gauge sector of the reduced theory. The analysis will we performed at the low energy regime. In this case the higher order derivatives terms in the potential of the HL Hamiltonian are dismissed and we are left with the electromagnetic coupling to HL gravity in the standard way. Despite the higher non--linear behavior of the resulting equation, we have thoroughly demonstrated the existence of a global minimum describing the existence of a throat connecting to regions, one of them asymptotically flat while the other presenting an essential singularity. Besides, the asymptotic behavior has been studied finding out that the solution tends to the well--known Reissner--Nordstr\"{o}m solution in GR. To support the feasibility of the present study we have incorporated a graphical analysis, where the assigned values to the coupling constants $k$, $\alpha$ and $\beta$ satisfy the known restrictions arising from experimental data. In fact, $\alpha$ and $\beta$ must be very near to 0 and 1 respectively \cite{feno5,feno6}. Consequently, the parameter $k$ we introduce satisfies
$k^{2}= 1- \alpha/2\beta >0$, $\beta >0$. 

The article is organized as follows: In Sec. \ref{sec2} are presented the 3+1 dimensional reduced theory and its equations of motion. In Sec. \ref{sec3} the main equations to be solved for the spherical charged throat are provided. Next, in Sec. \ref{sec4} the resulting differential equation is solved, showing the existence of the throat. Sec. \ref{sec5} shows the asymptotic behavior of the obtained solution and in Sec. \ref{sec6} the full metric charts of the charged throat are presented. Finally, Sec. \ref{sec7} concludes the work.

%%%%%%%%%%%%%%%%%%%%%%%%%%%%%%%%%%%%%%%%%%%%%%%%%%%
\section{The Anisotropic Gravity--Vector Gauge Interaction in the Ho\v{r}ava-Lifshitz Framework}\label{sec2}

The Hamiltonian formulation of the pure electromagnetic--gravitational interaction in the HL scenario was obtained and studied in \cite{r29,r30}. In that work the 3+1 Hamiltonian was obtained from a Kaluza--Klein reduction from a 4+1 dimensional HL gravity. At this point, we can take the dilaton scalar field be 1 in the Hamiltonian as one does in GR when following a Kaluza--Klein reduction to obtain the Einstein--Maxwell Hamiltonian. Otherwise one can leave the dilaton field unconstrained and obtained a formulation not equivalent to the previous one. In this case it is not possible to take the dilaton to be one in the field equations without restricting the other fields of the model. Both models are consistent and compatible with the geometric symmetries of the HL proposal. When the dilaton is taken to be 1, that is at its ground state, the Hamiltonian density describing this interaction is expressed as follows
\begin{equation}\label{eq1}
\begin{split}
\mathcal{H}= \frac{N}{\sqrt{\gamma}}\bigg[p^{ij}p_{ij}+\frac{p^{i}p_{i}}{2}+\frac{\lambda}{\left(1-4\lambda\right)}\left(p^{ij}\gamma_{ij}\right)^{2}-\gamma\beta\left(R-\frac{1}{4}F_{ij}F^{ij}\right)-\gamma\alpha a_{i}a^{i}\bigg] &\\
-\Lambda \partial_{i}p^{i} -\Lambda_{j}\left(2\nabla_{i}p^{ij}+p^{i}\gamma^{jk}F_{ik}\right)-\sigma P_{N} +\tilde{V}\left(\gamma_{ij},A_{k},N\right),
\end{split}
\end{equation}
where $\lambda\neq 1/4$. In the above expression $\tilde{V}\left(\gamma_{ij},A_{k},N\right)$ represents the higher order spatial derivatives terms in the potential of the theory
and, $R$ is the curvature associated with the $3$-dimensional metric $\gamma_{ij}$. Furthermore, $\Lambda$, $\Lambda_{j}$ and $\sigma$ Lagrange multipliers and the two form $F_{ij}dx_{i} \wedge dx_{j}$ is defined by
\begin{equation}\label{eq2}
F_{ij}=\partial_{i}A_{j}-\partial_{j}A_{i}.   
\end{equation}
From now on we will dismiss the higher order derivative terms contained in the potential $\tilde{V}$. Physically, it corresponds to consider the low energy Hamiltonian. So, Following \cite{r29}, the first class constraints associated to (\ref{eq1}) are given by 
\begin{eqnarray}\label{eq3}
\partial_{i}p^{i}&=&0 \\ \label{eq4}
 2\nabla_{i}p^{ij}+p^{i}\gamma^{jk}F_{ik}&=&0,
\end{eqnarray}
and the second class one is
\begin{equation}\label{eq5}
\begin{split}
H_{N}\equiv\frac{1}{\sqrt{\gamma}}\bigg[p^{ij}p_{ij}+\frac{p^{i}p_{i}}{2}+\frac{\lambda}{\left(1-4\lambda\right)}\left(p^{ij}\gamma_{ij}\right)^{2}-\beta\gamma R +\frac{\beta}{4} \gamma F_{ij}F^{ij} \bigg]+\alpha\sqrt{\gamma}a_{i}a^{i}\\
+2\alpha\sqrt{\gamma}\nabla_{i}a^{i}
=0.
\end{split}
\end{equation}
Next, the Hamiltonian density (\ref{eq1}) provides the following evolution equations
\begin{eqnarray}\label{eq6}
\dot{\gamma}_{ij}&=&\frac{2N}{\sqrt{\gamma}}\left[p_{ij}+\frac{\gamma_{ij}\lambda}{\left(1-4\lambda\right)}p^{lm}\gamma_{lm}\right]+\nabla_{i}\Lambda_{j}+\nabla_{j}\Lambda_{i}, \\ \label{eq7}
\dot{A}_{i}&=&\frac{N}{\sqrt{\gamma}}p_{i}+\partial_{i}\Lambda-\Lambda_{j}\gamma^{jk}F_{ik},
\end{eqnarray}
\begin{equation}\label{eq8}
\begin{split}
\dot{p}^{ij}=\frac{N}{2}\frac{\gamma^{ij}}{\sqrt{\gamma}}\left[p^{lk}p_{lk}+\frac{p^{l}p_{l}}{2}+\frac{\lambda}{\left(1-4\lambda\right)}\left(p^{lm}\gamma_{lm}\right)^{2}\right]  
-\frac{N}{\sqrt{\gamma}}\bigg[2p^{ik}p^{j}_{\ k}+\frac{p^{i}p^{j}}{2} &\\+\frac{2\lambda}{\left(1-4\lambda\right)}\left(p^{lm}\gamma_{lm}\right)p^{ij}\bigg]
+N\sqrt{\gamma}\beta\left[\frac{R}{2}\gamma^{ij}
- R^{ij}\right]+\beta\sqrt{\gamma}\bigg[\nabla^{(i}\nabla^{j)}N &\\-\gamma^{ij}\nabla_{k}\nabla^{k}N\bigg]
+\frac{\beta}{2}N\sqrt{\gamma}\Bigg[F^{in}F^{\ j}_{n}
-\frac{\gamma^{ij}}{4}F_{mn}F^{mn}\Bigg]
+\alpha N\sqrt{\gamma} \bigg[\frac{\gamma^{ij}}{2}a_{k}a^{k}
&\\-a^{i}a^{j}\bigg]
-\nabla_{k}\bigg[2p^{k(i}\Lambda^{j)}-p^{ij}\Lambda^{k}\bigg]-\Lambda^{(i}\gamma^{j)m}p^{l}F_{lm},
\end{split}    
\end{equation}
\begin{equation}\label{eq9}
\dot{p}^{i}=\beta\partial_{j}\left(N\sqrt{\gamma} F^{ji}\right)+\partial_{k}\left(\Lambda^{k}p^{i}-\Lambda^{i}p^{k}\right).
\end{equation}
\section{The Spherically Symmetric and Static Equations}\label{sec3}
As we are interested in the study of spherically symmetric, static and charged configurations we consider the following line element describing the 3--dimensional manifold
\begin{equation}\label{eq10}
ds^{2}_{(3)}=\frac{dr^{2}}{f(r)}+r^{2}d\Omega^{2},
\end{equation}
together with a lapse function $N=N(r)$ and a shift $\Lambda_{i}=0$, where $d\Omega^{2}\equiv\ d\theta^{2}+\text{sin}^{2}\theta d\varphi^{2}$. Besides, the staticity assumption implies that the magnetic field is null \i.e, $F_{ij}=0$, $\dot{A}_{i}=0$ and $\dot{\gamma}_{ij}=0$ then from Eq. (\ref{eq6}) one obtains $p^{ij}=0$ for $\lambda$ different from 1 and $1/4$. The latter means that the terms containing the dimensionless constant $\lambda$ are canceled on the field equations. At this point it is worth mentioning that the static electric field is encoded in $p^{i}$. The field equations (\ref{eq7})--(\ref{eq8}) reduce then to 
\begin{equation}\label{eq11}
\frac{N}{\sqrt{\gamma}}p_{i}+\partial_{i}\Lambda=0,
\end{equation}
\begin{equation}\label{eq12}
\begin{split}
\frac{1}{4}\frac{\gamma^{ij}}{\gamma}p^{k}p_{k}-\frac{1}{2\gamma}p^{i}p^{j}+\beta\left(\frac{\gamma^{ij}}{2}R-R^{ij}\right)+\frac{\beta}{N}\left(\nabla^{(i}\nabla^{j)}N-\gamma^{ij}\nabla_{k}\nabla^{k}N\right) &\\
+\alpha\left(\frac{\gamma^{ij}}{2}a_{k}a^{k}-a^{i}a^{j}\right)=0.
\end{split}
\end{equation}
Now, taking the trace of the above expression (\ref{eq12}) one gets
\begin{equation}\label{eq13}
\frac{1}{4\gamma}p^{k}p_{k}+\frac{1}{2}\beta R-2\frac{\beta}{N}\nabla_{i}\nabla^{i}N+\frac{\alpha}{2}a_{k}a^{k}=0.    
\end{equation}
From Eq. (\ref{eq5}) one arrives to
\begin{equation}\label{eq14}
\frac{1}{2}\frac{1}{\gamma}p^{k}p_{k}-\beta R-\alpha a^{i}a_{i} +2\alpha\frac{\nabla_{i}\nabla^{i}N}{N}=0.    
\end{equation}
Combining Eqs. (\ref{eq13}) and (\ref{eq14}) one gets
\begin{equation}\label{eq15}
\frac{1}{2}p^{k}p_{k}+\left(\alpha-2 \beta\right)\frac{\nabla_{i}\nabla^{i}N}{N}=0.    
\end{equation}
Due to the spherical symmetry the only non vanishing component in (\ref{eq11}) is the radial one. So, we have
\begin{equation}\label{eq16}
\frac{p_{r}}{\sqrt{\gamma}}=-\frac{\partial_{r}\Lambda(r)}{N}.    
\end{equation}
Raising the indices with the $3$-dimensional metric tensor $\gamma^{ij}$ from (\ref{eq16}) one gets
\begin{equation}\label{eq17}
\frac{p^{r}}{\sqrt{\gamma}}=-\frac{f(r)}{N}\partial_{r}\Lambda(r).     
\end{equation}
Now from (\ref{eq3}) and (\ref{eq11}) we get
\begin{equation}\label{eq19}
\frac{r^{2}}{N}\sqrt{f}\partial_{r}\Lambda=A,    
\end{equation}
being $A$ a new integration constant. We then have for the static electric field 
\begin{equation}\label{eq20}
\sqrt{\frac{p_{r}p^{r}}{\gamma}}=\frac{|A|}{r^{2}}.
\end{equation}
Next, putting together Eqs. (\ref{eq15}), (\ref{eq16}), (\ref{eq17}) and (\ref{eq19}) we arrive at the following equation
\begin{equation}\label{eq18}
\frac{1}{4}\partial_{r}\Lambda^{2}+\frac{1}{2}\left(\alpha-2\beta\right)\partial_{r}N^{2}=\frac{CN}{r^{2}\sqrt{f}},    
\end{equation}
where $C$ is an integration constant. By replacing (\ref{eq19}) on the right hand side of (\ref{eq18}) we obtain
\begin{equation}\label{eq21}
\frac{\Lambda^{2}}{4}+\frac{1}{2}\left(\alpha-2\beta\right)N^{2}=\frac{C}{A}\Lambda +B,    
\end{equation}
where $B$ is an integration constant. Eq. (\ref{eq21}) can be reorganized as follows 
\begin{equation}\label{eq22}
\left(\frac{\alpha}{2}-\beta\right)N^{2}+\left(\frac{\Lambda}{2}-\frac{C}{A}\right)^{2}=\frac{C^{2}}{A^{2}}+B\equiv E.  
\end{equation}
Equation (\ref{eq12}) has only two independent components, the $r-r$ and the $\theta-\theta$ components. The complete set of field equations then reduces to the $\theta-\theta$ component of Eq. (\ref{eq12}), its trace (\ref{eq13}), Eq. (\ref{eq15}) or equivalently Eqs. (\ref{eq22}), (\ref{eq17}) and (\ref{eq19}). 

For the $3$-dimensional metric under consideration we have the following components of the Ricci tensor
\begin{equation}\label{eq23}
R^{rr}=-\frac{f}{r}\partial_{r}f, \quad R^{\theta\theta}=-\frac{r\partial_{r}f+2f-2}{2r^{4}} \quad \mbox{and} \quad R^{\varphi\varphi}=csc^{2}(\theta)R^{\theta\theta}, 
\end{equation}
and the Ricci's scalar is given by
\begin{equation}\label{eq24}
R=-\frac{2}{r^{2}}\left(f-1+r\partial_{r}f\right).  
\end{equation}
So, the $\theta-\theta$ component of Eq. (\ref{eq12}), after the replacement of (\ref{eq13}) on it, is
\begin{equation}\label{eq25}
\frac{\beta}{2r^{4}}\left(r\partial_{r}f+2f-2\right)+\frac{\beta}{Nr^{3}}f\partial_{r}N+\frac{\beta}{r^{2}N}\nabla_{i}\nabla^{i}N=0,    
\end{equation}
while Eq. (\ref{eq13}), after the replacement of Eqs. (\ref{eq16}), (\ref{eq17}) and (\ref{eq19}) on it, is
\begin{equation}\label{eq26}
-\frac{\beta}{r^{2}}\left(r\partial_{r}f+f-1\right)+\frac{\alpha}{2}f\left(\frac{\partial_{r}N}{N}\right)^{2}-2\frac{\beta}{N}\nabla_{i}\nabla^{i}N=-\frac{A^{2}}{4r^{4}}.   
\end{equation}
Finally, from Eqs. (\ref{eq25}) and (\ref{eq26}) we get
\begin{equation}\label{eq27}
\left(\sqrt{f}+r\sqrt{f} \frac{\partial_{r}N}{N}\right)^{2}-\left(r\sqrt{1-\frac{\alpha}{2\beta}}\sqrt{f}\frac{\partial_{r}N}{N}\right)^{2}=1-\frac{A^{2}}{4\beta r^{2}}.    
\end{equation}
Consequently, the independent field equations reduce to (\ref{eq17}), (\ref{eq19}), (\ref{eq22}), (\ref{eq25}) or (\ref{eq26}) and (\ref{eq27}). 
\section{The Charged Throat Solution}\label{sec4}
We consider a non--trivial electrostatic field, hence $A\neq 0$. We will now introduce a new variable $\chi$. This new variable will be the correct one to describe the metric on the throat. Eq. (\ref{eq27}) can be solved in terms of the new variable $\chi$. From now on we shall employ prime to denotes derivatives with respect to the radial coordinate $r$,
\begin{equation}\label{eq28}
kr\sqrt{f}\frac{N^{\prime}}{N}=\pm\sqrt{1-\frac{A^{2}}{4\beta r^{2}}}\text{Sinh}(\chi),   
\end{equation}
\begin{equation}\label{eq29}
\sqrt{f}\left(1+r\frac{N^{\prime}}{N}\right)=\pm\sqrt{1-\frac{A^{2}}{4\beta r^{2}}}\text{Cosh}(\chi).  
\end{equation}
It then follows
\begin{equation}\label{eq30}
\sqrt{f}=\pm\sqrt{1-\frac{A^{2}}{4\beta r^{2}}}\left(\text{Cosh}(\chi)-\frac{1}{k}\text{Sinh}(\chi)\right), 
\end{equation}
in (\ref{eq28})--(\ref{eq30}) we take + for $\text{Cosh}(\chi) \geq\frac{1}{k} \text{Sinh}(\chi)$ and - for $\text{Cosh}(\chi)< \frac{1}{k}\text{Sinh}(\chi)$, and
\begin{equation}\label{eq31}
\frac{N^{\prime}}{N}=\frac{\text{Sinh}(\chi)}{r\left(k\text{Cosh}(\chi)-\text{Sinh}(\chi)\right)}.   
\end{equation}
From (\ref{eq20}) and (\ref{eq22}) we have
\begin{equation}\label{eq32}
\Lambda=2\frac{C}{A}+2\sqrt{E-\left(\frac{\alpha}{2}-\beta\right)N^{2}},    
\end{equation}
\begin{equation}\label{eq33}
\frac{\Lambda^{\prime}}{N}=2\beta k^{2}\frac{N^{\prime}}{\sqrt{E-\left(\frac{\alpha}{2}-\beta\right)N^{2}}}=\frac{A}{r^{2}\sqrt{f}}.
\end{equation}
Now using (\ref{eq28}) we obtain
\begin{equation}\label{eq34}
N^{2}=\frac{EA^{2}}{\left(2k\beta\right)^{2}\left(r^{2}\text{Sinh}^{2}(\chi)-\frac{A^{2}}{4\beta}\text{Cosh}^{2}(\chi)\right)},   
\end{equation}
we will show that its right hand member is positive. From (\ref{eq31}) and (\ref{eq34}) we finally obtain a first order differential equation determining $r$ as a function of $\chi$. Explicitly it reads
\begin{equation}\label{eq35}
\begin{split}
\frac{dr}{d\chi}=-\left(\frac{\text{Cosh}(\chi)}{\text{Sinh}(\chi)}-\frac{1}{k}\right)r\left(1-\frac{A^{2}}{4\beta r^{2}}\right)\left(1-\frac{A^{2}}{4k\beta r^{2}}\frac{\text{Cosh}(\chi)}{\text{Sinh}(\chi)}\right)^{-1}.   
\end{split}    
\end{equation}
In the particular case in which $A=0$ this differential equation was solved explicitly in \cite{r31}. Moreover, the expressions for $\sqrt{f}$ in (\ref{eq30}) and $N^{2}$ in (\ref{eq34}) reduces to the ones in that reference. In order to analyze the first order differential equation (\ref{eq35}) and following the theorem of existence and uniqueness of the solution to ordinary differential equations we give the initial data. At $\hat{\chi}$ we give $\hat{r}$ defined as follows: $\hat{\chi}$ is the solution of the equation
\begin{equation}\label{eq36}
\frac{\text{Cosh}(\hat{\chi})}{\text{Sinh}(\hat{\chi})}=\frac{1}{k},    
\end{equation}
and we take $\hat{r}$ any real number satisfying 
\begin{equation}\label{eq37}
\hat{r}>\frac{|A|}{2k\sqrt{\beta}}L
\end{equation}
We have already assumed $k^{2}=1-\frac{\alpha}{2\beta}>0$, $\beta>0$ in agreement with the experimental data. We assume now, $\alpha>0$, hence we have $k<1$. There exists then the solution $\hat{\chi}$ to Eq. (\ref{eq36}). $L$ is defined as
$L= (1/k) [ (1+k)^{1+k}/ (1-k)^{1-k}]^{1/2k}=e^{(\hat{\chi}/k)}/\text{Sinh}(\hat{\chi})$,
it is $L>1$ and when $k$ goes to 1, then $L$ tends to 2.
 We notice that if $\alpha=0$ there does not exist $\hat{\chi}$, in this case the field equations on the previous section, after a redefinition in order to eliminate $\beta$, are the corresponding ones in GR. We then have at $\hat{\chi}$
\begin{equation}\label{eq38}
\frac{dr}{d\chi}\bigg{|}_{\hat{\chi}}=0,
\end{equation}
\begin{equation}\label{eq39}
\frac{d^{2}r}{d\chi^{2}}\bigg{|}_{\hat{\chi}}=\frac{\hat{r}}{\left(\text{Sinh}\hat{\chi}\right)^{2}}\left(1-\frac{A^{2}}{4\beta \hat{r}^{2}}\right)\left(1-\frac{A^{2}}{4k^{2}\beta \hat{r}^{2}}\right)^{-1}>0,    
\end{equation}
hence we have a local minimum at $\hat{\chi}$. We will show now that it is a global minimum of $r=r(\chi)$.\\

Let us denote $h(\chi,r(\chi))$ the right hand member of Eq. (\ref{eq39}). Since at $\hat{\chi}$, $r(\hat{\chi})=\hat{r}$ satisfies the bound (\ref{eq37}), then by continuity of the factors in $h(\chi,r(\chi))$ there exists a neighborhood $U$ of $\hat{\chi}$ such that
\begin{equation}\label{eq40}
1-\frac{A^{2}}{4\beta r^{2}}>0, \quad 1-\frac{A^{2}}{4k^{2}\beta {r}^{2}}>0,    
\end{equation}
for all $\chi \in U$. Let us denote $U_{+}\equiv \{\chi \in U: \chi>\hat{\chi}\}$ and $U_{-}\equiv \{\chi \in U: \chi<\hat{\chi}\}$.\\

Then for all $\chi \in U_{+}$ 
\begin{equation}\label{eq41}
\frac{\text{Cosh}({\chi})}{\text{Sinh}({\chi})}<\frac{1}{k},    
\end{equation}
hence $h(\chi,r(\chi))>0$ and $dr/d\chi>0$. Consequently, $r(\chi)$ is a monotonically increasing function of $\chi$. Then 
\begin{equation}\label{eq42}
\frac{dr}{d\chi}>\left(\frac{1}{k}-\frac{\text{Cosh}(\chi)}{\text{Sinh}(\chi)}\right)\left(1-\frac{A^{2}}{4\beta r^{2}}\right)r>0,
\end{equation}
where the right hand member is a monotonically increasing function of $\chi$, for all $\chi \in U_{+}$. But now we can extend the neighborhood $U_{+}$, for all  $\chi > \hat{\chi}$ and (\ref{eq40}), (\ref{eq41}) and (\ref{eq42}) remain valid for the extension, the function $r(\chi)$ is then a monotonically increasing function for all $\chi> \hat{\chi}$, with an infimum value at  $\hat{r}$.
We can also bound $dr/d\chi$ from above for all $\chi>\hat{\chi}$. In fact
\begin{equation}\label{eq43}
0<\frac{dr}{d\chi}<\left(\frac{1}{k}-\frac{\text{Cosh}(\chi)}{\text{Sinh}(\chi)}\right)r\left(1-\frac{A^{2}}{4\beta r^{2}}\right)\left(1-\frac{A^{2}}{4k^{2}\beta r^{2}}\right)^{-1}.    
\end{equation}
Hence $dr/d\chi$ is bounded from above and from below by strictly positive differentiable functions of $\chi$. The solution $r(\chi)$ extends then as a differentiable function for all $\chi>\hat{\chi}$ and $r(\chi)\rightarrow+\infty$ when $\chi\rightarrow +\infty$. Indeed, $\frac{dr}{d\chi} \sim \left(\frac{1}{k}-1\right)r$ when $\chi\rightarrow +\infty$.\\

\begin{figure}[H]
\centering
\includegraphics[width=0.32\textwidth]{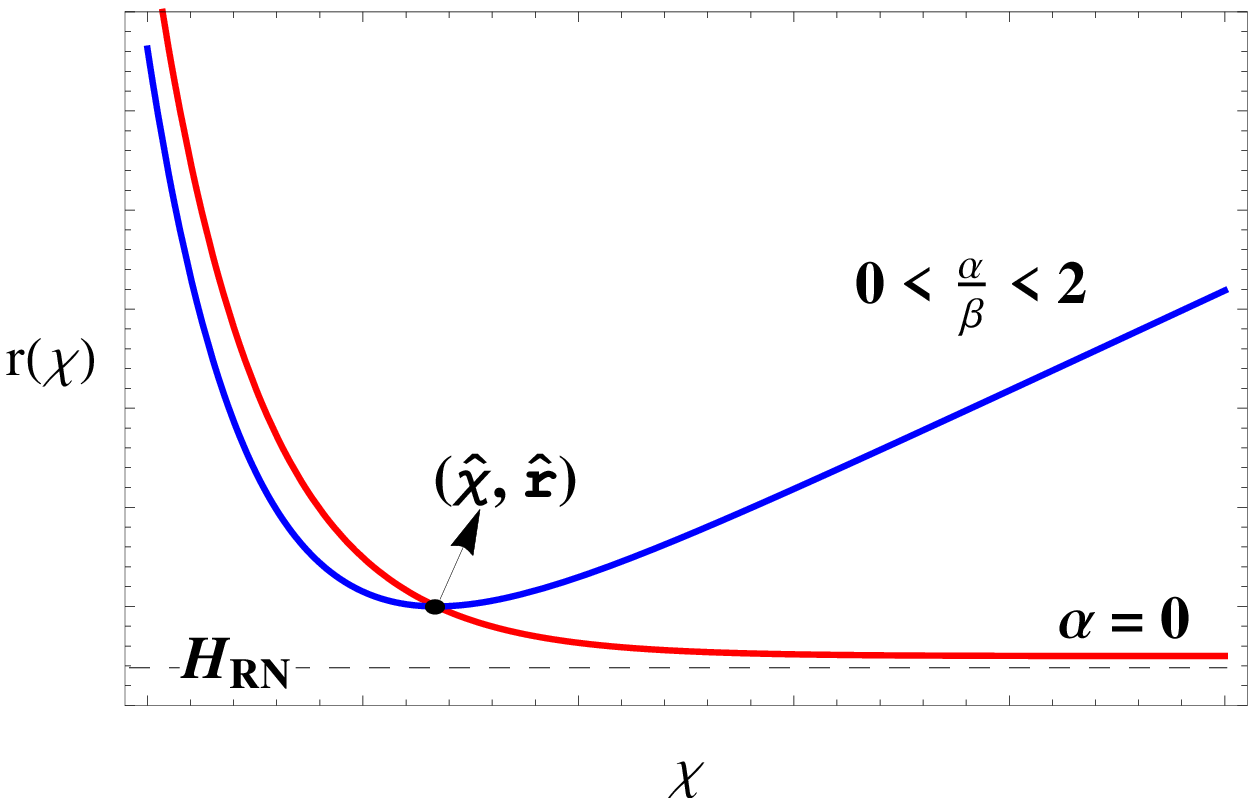}
\includegraphics[width=0.32\textwidth]{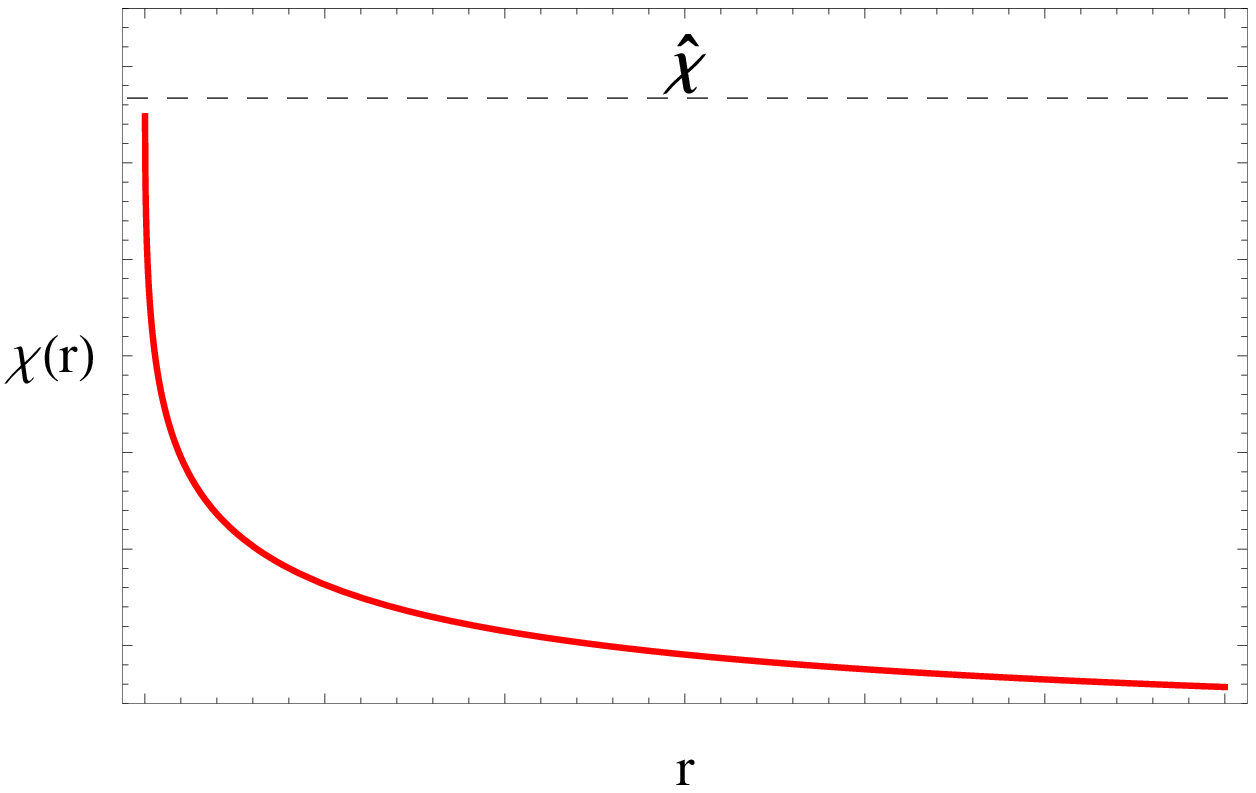}
\includegraphics[width=0.32\textwidth]{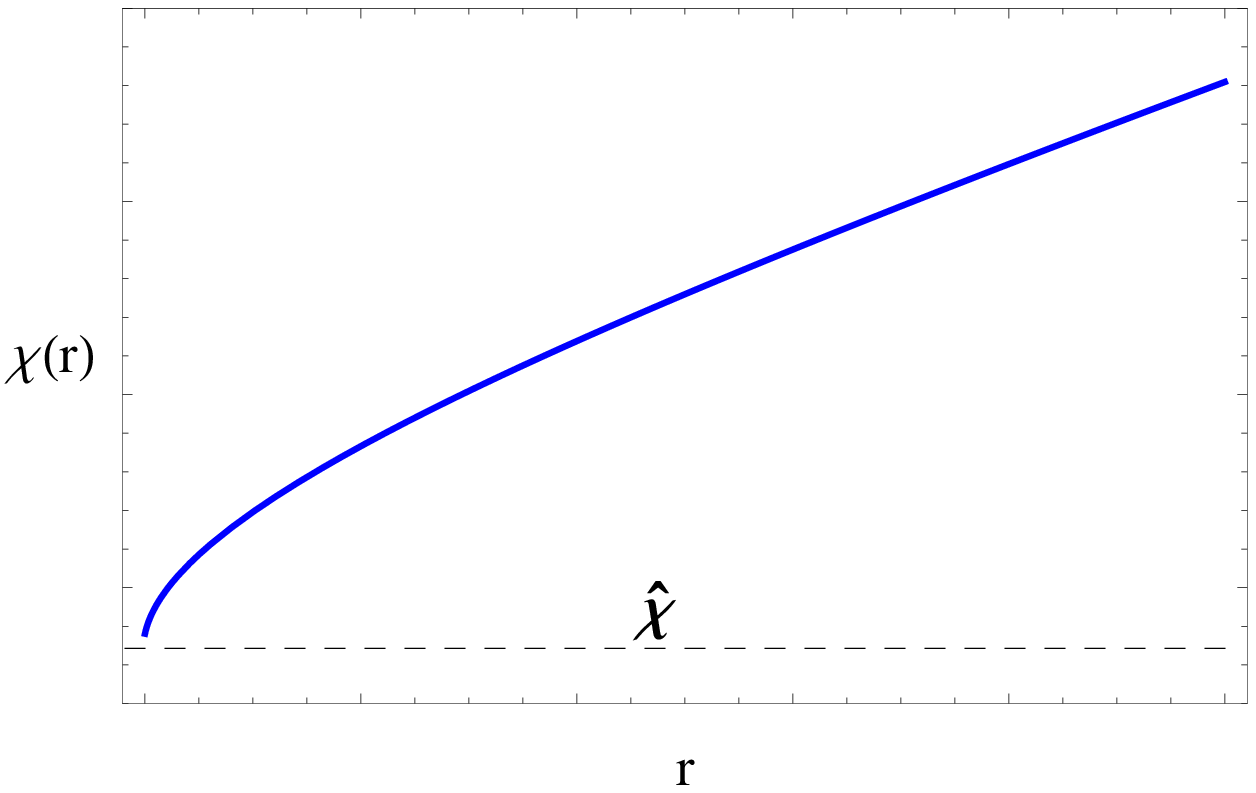}
\caption{\textbf{Left panel:} The throat located at the point $(\hat{\chi},\hat{r})$ (blue curve) and the GR limit (red curve), against $\chi$. \textbf{Middle panel:} The $\chi$ variable as a function of the radial $r$ one, for points $\chi<\hat{\chi}$. \textbf{Right panel:} The $\chi$ variable versus the radial $r$ one for points $\chi>\hat{\chi}$.
}
\label{fig1}
\end{figure}

We may now analyze the solution for $\chi<\hat{\chi}$. We consider $\chi \in U_{-}$ and introduce
\begin{equation}\label{eq44}
\omega(\chi)=\chi-\frac{1}{k}\text{Ln}\left[\sqrt{1-\frac{A^{2}}{4\beta r^{2}}}\right],    
\end{equation}
then 
\begin{equation}\label{eq45}
\frac{d\omega}{d\chi}=1-\frac{A^{2}}{4rk\beta\left(r^{2}-\frac{A^{2}}{4\beta}\right)}\frac{dr}{d\chi}.    
\end{equation}
So, we obtain 
\begin{equation}\label{eq46}
\frac{dr}{d\chi}=\frac{d\omega}{d\chi}\left(\frac{1}{k}-\frac{\text{Cosh}(\chi)}{\text{Sinh}(\chi)}\right)r\left(1-\frac{A^{2}}{4\beta r^{2}}\right)\left(1-\frac{A^{2}}{4k^{2}\beta r^{2}}\right)^{-1}.    
\end{equation}
We notice that 
\begin{equation}\label{eq47}
\frac{1}{k}-\frac{\text{Cosh}(\hat{\chi})}{\text{Sinh}(\hat{\chi})}<0,    
\end{equation}
for all $\chi \in U_{-}$ and it is a monotonically increasing function of $\chi$. From the theorem of existence and uniqueness of the solution to the differential equation (\ref{eq35}) we conclude that there exists $r(\chi)$ satisfying the initial data $r(\hat{\chi})=\hat{r}$ at least in a neighborhood $U_{-}$ of $\hat{\chi}$, and it is differentiable. We have for $\chi \in U_{-}$: $\frac{dr}{d\chi}<0$. Hence $\frac{d\omega}{d\chi}>1$. We then obtain the following bound for $\frac{dr}{d\chi}$
\begin{equation}\label{eq48}
-\frac{dr}{d\chi}>-\left(\frac{1}{k}-\frac{\text{Cosh}(\chi)}{\text{Sinh}(\chi)}\right)r>0,    
\end{equation}
for all $\chi \in U_{-}$. We can integrate both side of (\ref{eq48}) from $\chi$ to $\hat{\chi}$, obtaining 
\begin{equation}\label{eq49}
r>\hat{r}\frac{e^{\chi/k}}{\text{Sinh}(\chi)}\frac{\text{Sinh}(\hat{\chi})}{e^{\hat{\chi}/k}}=\frac{\hat{r}}{L}\frac{e^{\chi/k}}{\text{Sinh}(\chi)},  
\end{equation}
and also
\begin{equation}\label{eq50}
1>1-\frac{A^{2}}{4k\beta r^{2}}\frac{\text{Cosh}(\chi)}{\text{Sinh}(\chi)}>1- \frac{A^{2}}{4k\beta}\frac{L^{2}}{\hat{r}^{2}}>0,
\end{equation}
where we have used (\ref{eq49}) and (\ref{eq37}). Additionally from equations (\ref{eq35}) and (\ref{eq50}) we get
\begin{equation}\label{eq51}
-\frac{1}{r}\frac{dr}{d\chi}<\left(\frac{\text{Cosh}(\chi)}{\text{Sinh}(\chi)}-\frac{1}{k}\right)\left(1-\frac{A^{2}L^{2}}{4\beta k\hat{r}^{2}}\right)^{-1},   
\end{equation}
which implies
\begin{equation}\label{eq52}
r<\frac{\hat{r}}{L}\frac{e^{\chi/k}}{\text{Sinh}(\chi)}\left(1-\frac{A^{2}L^{2}}{4\beta k\hat{r}^{2}}\right)^{-1}.   
\end{equation}
The differential function $r(\chi)$ is a monotonically decreasing function , bounded, from below by (\ref{eq49}) and from above by (\ref{eq52}), the function can then be extended for all $\chi<\hat{\chi}$ with the same  properties and bounds. Its infimum values is obtained at $\hat{r}=r(\hat{\chi})$ and when $\chi\rightarrow0^{+}$ then $r(\chi)\rightarrow+\infty$. We conclude then that $r(\chi)$ is a differentiable function with a global minimum at $\hat{\chi}$, it is monotonically decreasing for $0<\chi<\hat{\chi}$ and monotonically increasing for $\hat{\chi}<\chi$. Also, $r\rightarrow+\infty$ when $\chi\rightarrow0^{+}$ and $r\rightarrow+\infty$ when $\chi\rightarrow+\infty$. The previous analysis is depicted in Fig. \ref{fig1}, where the left panel shows the existence of the global minimum $\hat{r}=r(\hat{\chi})$ describing the throat (blue curve). To compare the resulting solution with the GR limit, we have plotted the solution of (\ref{eq35}) it for the values $\{\beta;\alpha;k\}=\{1;0;1\}$ represented by the red curve. As it is appreciated, for the GR limit there isn't a throat. Besides, the middle and right panels are displaying the behavior of the $\chi$ coordinate in terms of the radial $r$ one. As was mentioned , it is clear the one to one correspondence between both variables away from the throat. Moreover, the radial coordinate $r$ is divergent when $\chi\rightarrow0^{+}$ or $\chi\rightarrow+\infty$, the $\chi$ is the correct one to describe the charged throat.

\section{Asymptotic Behaviour of the Solutions}\label{sec5}

In this section we analyze the asymptotic behavior of the 
solution. Interestingly, this analysis can be done either, starting from the original equations without the introduction of the variable $\chi$ or from the previous  analysis in terms of $\chi$. So, from (\ref{eq15}) and (\ref{eq20}) we obtain
\begin{equation}\label{eq53}
\frac{\nabla_{i}\nabla^{i}N}{N}=\frac{A^{2}}{4k^{2}r^{4}\beta}.    
\end{equation}
We replace it on (\ref{eq25}) to get
\begin{equation}\label{eq54}
\frac{r}{2}f'+f-1+rf\frac{N'}{N}=-\frac{A^{2}}{4k^{2}r^{2}\beta}.    
\end{equation}
From (\ref{eq26}) we have 
\begin{equation}\label{eq55}
\frac{1}{r^{2}}\left(rf'+f-1\right)-\frac{\alpha}{2\beta}f\left(\frac{N'}{N}\right)^{2}=\frac{A^{2}}{2\beta r^{4}}\left(\frac{1}{2}-\frac{1}{k^{2}}\right).
\end{equation}
We notice from (\ref{eq30}), that for $\chi<\hat{\chi}$, $\sqrt{f}$ is bounded when $r\rightarrow +\infty$. We can then expand $f$, asymptotically when $r\rightarrow +\infty$, as  
\begin{equation}\label{eq56}
f=f_{0}+\frac{a}{r}+\frac{b}{r^{2}}+\mathcal{O}\left(\frac{1}{r^{3}}\right).    
\end{equation}
After replacing this expression for $f$ in the previous equations we obtain, assuming $a\neq0$,
\begin{equation}\label{eq57}
f_{0}=1,    
\end{equation}
and
\begin{equation}\label{eq58}
b=-\frac{\alpha}{2\beta}\frac{a^{2}}{4}+\frac{A^{2}}{2\beta}\left(\frac{1}{k^{2}}-\frac{1}{2}\right). 
\end{equation}
In the particular case $\alpha=0$, $\beta=1$ we obtain $b=\frac{A^{2}}{4}$, $f$ coincides in this case with the Reissner-Nordstr\"om solution in GR. From the above equations we also obtain the lapse $N$ as follows
\begin{equation}\label{eq59}
N^{2}=N^{2}_{0}\left[1+\frac{a}{r}+\frac{A^{2}}{4k^{2}r^{2}\beta}+\mathcal{O}\left(\frac{1}{r^{3}}\right)\right].    
\end{equation}
where we shall take $N_{0}=1$. As before, if $\alpha=0$ and $\beta=1$ we then obtain the same asymptotic behaviour of the metric as in the Reissner--Nordstr\"om space-time in GR. In order to express the parameter $a$ in terms of the initial data $\{\hat{r},\hat{\chi}\}$, we consider the asymptotic expansion of $f$ from the expansion (\ref{eq30}). Therefore, we take
\begin{equation}\label{eq60}
\chi=\frac{m}{r}+\frac{n}{r^{2}}+\mathcal{O}(\frac{1}{r^{3}}).    
\end{equation}
From the differential equation (\ref{eq35}) we obtain
\begin{equation}\label{eq61}
n=\frac{m^{2}}{k}-\frac{A^{2}}{4\beta k},    
\end{equation}
and from the expansion of (\ref{eq30}) we get
\begin{equation}\label{eq62}
a=-\frac{2}{k}m,   
\end{equation}
that is 
\begin{equation}\label{eq63}
a=-\frac{2}{k}\lim_{\chi\rightarrow0^{+}}r(\chi)\text{Sinh}(\chi).    
\end{equation}
The above limit exits and it is bounded by the following expressions
\begin{equation}\label{eq64}
\frac{\hat{r}}{L}\leq\lim_{\chi\rightarrow0^{+}}r(\chi)\text{Sinh}(\chi)\leq \frac{\hat{r}}{L}\left(1-\frac{A^{2}L^{2}}{4\beta k \hat{r}^{2}}\right)^{-1}.     
\end{equation}

\section{The Metric of the Charged Throat}\label{sec6}
We use two coordinate charts to describe the 4--dimensional metric describing the charged throat solution of the electromagnetic--gravitational HL field equations. In one chart the radial coordinate $r$ is used. In that chart defined as $r> r(\hat{\chi}- \epsilon)$ that is, $r$ evaluated at $\hat{\chi}- \epsilon$ or equivalently $0<\chi<\hat{\chi}-\epsilon$, with $\epsilon>0$, the metric is given by
\begin{equation}\label{65}
ds^{2}=-N^{2}(r)dt^{2}+\frac{dr^{2}}{f(r)}+r^{2}d\Omega^{2},    
\end{equation}
where $f(r)$ is given by
\begin{equation}\label{eq66}
f(r)=\left(1-\frac{A^{2}}{4r^{2}\beta}\right)\left(\text{Cosh}\chi(r)-\frac{1}{k}\text{Sinh}\chi(r)\right)^{2},  
\end{equation}
and $N^{2}$ by (\ref{eq34}).\
When $0<\chi<\hat{\chi}$ we have shown that the solution to the field equations satisfies $r>\hat{r}$, where $\hat{r}$ is the value of $r$ at the throat. It is the infimum of $r$ on that interval. Hence
\begin{equation}\label{eq67}
f(r)>\left(1-\frac{A^{2}}{4\hat{r}^{2}\beta}\right)\left(\text{Cosh}(\hat{\chi}-\epsilon)-\frac{1}{k}\text{Sinh}(\hat{\chi}-\epsilon)\right)^{2}>0, \quad 0<\chi<\hat{\chi}-\epsilon  
\end{equation}

and $f(r)\rightarrow 1$ when $r\rightarrow +\infty$.\  
The function $\chi(r)$ is the solution to 
\begin{equation}\label{eq68}
-\frac{d\chi}{dr}=\frac{1}{r}\left(\frac{\text{Cosh}(\chi)}{\text{Sinh}(\chi)}-\frac{1}{k}\right)^{-1}\left(1-\frac{A^{2}}{4\beta r^{2}}\right)^{-1}\left(1-\frac{A^{2}}{4k\beta r^{2}}\frac{\text{Cosh}(\chi)}{\text{Sinh}(\chi)}\right),  
\end{equation}
we have shown that the solution $\chi(r)$ or equivalently $r(\chi)$ exist, they are differentiable functions and there is a one to one correspondence between $\chi$ and $r$, moreover $\frac{dr}{d\chi}<0$ for $0<\chi<\hat{\chi}$. All the factors on the right hand side are positive on that range of $\chi$. We now show that $N^{2}$ is also well defined for $0<\chi<\hat{\chi}$. In fact, the denominator of the expression of $N^{2}$, Eq. (\ref{eq34}), is bounded from below by
\begin{equation}\label{eq69}
\left(r\text{Sinh}(\chi)\right)^{2}-\frac{A^{2}}{4\beta}\left(\text{Cosh}(\chi)\right)^{2}\geq \left(\frac{\hat{r}^{2}}{L^{2}}-\frac{A^{2}}{4\beta}\right)\left(\text{Cosh}(\chi)\right)^{2}\geq \frac{\hat{r}^{2}}{L^{2}}-\frac{A^{2}}{4\beta}>0.    
\end{equation}
Moreover, the $\lim r(\chi)\text{Sinh}(\chi)$ when $\chi\rightarrow0^{+}$ exits and it is bounded (\ref{eq64}). We then take the integration constant $E$ such that 
\begin{equation}\label{eq70}
\lim _{\chi\rightarrow 0^{+}}N^{2}=1.
\end{equation}

The second chart in order to describe the throat solution is defined in terms of $\chi$ for $\hat{\chi}-\epsilon<\chi$. The metric is given by 
\begin{equation}\label{eq71}
ds^{2}=-N^{2}(r(\chi))dt^{2}+\left(\frac{dr}{d\chi}\right)^{2}\frac{d\chi^{2}}{f(r(\chi))}+r^{2}(\chi)d\Omega^{2}, 
\end{equation}
where $r(\chi)$ is the solution of the differential equation (\ref{eq35}). It is a differentiable function with a minimum at $\hat{r}=r(\hat{\chi})$.. The expression 
\begin{equation}\label{eq72}
\left(\frac{dr}{d\chi}\right)^{2}\frac{1}{f(r(\chi))}=\frac{1}{\left(\text{Sinh}(\chi)\right)^{2}}\left(1-\frac{A^{2}}{4\beta r^{2}}\right)\left(1-\frac{A^{2}}{4\beta k r^{2}}\frac{\text{Cosh}(\chi)}{\text{Sinh}(\chi)}\right)^{-2},   
\end{equation}
is non--singular and bounded away from zero on the throat $\hat{\chi}-\epsilon<\chi<\hat{\chi}+\epsilon$. Indeed, 
\begin{equation}\label{eq73}
1-\frac{A^{2}}{4\beta r^{2}}\geq 1-\frac{A^{2}}{4\beta\hat{r}^{2}}>0 \quad \forall \quad \chi,   
\end{equation}
\begin{equation}\label{eq74}
1-\frac{A^{2}}{4\beta k r^{2}}\frac{\text{Cosh}(\chi)}{\text{Sinh}(\chi)}\geq1-\frac{A^{2}}{4\beta k^{2} \hat{r}^{2}}>0 \quad \forall \quad \chi\geq\hat{\chi},    
\end{equation}
and the bound (\ref{eq50}) for all $\chi<\hat{\chi}$. Furthermore, $N^{2}$ is well defined on the throat, since
\begin{equation}\label{eq75}
\left(r\text{Sinh}(\chi)\right)^{2} -\frac{A^{2}}{4\beta}\left(\text{Cosh}(\chi)\right)^{2}\geq\left(\hat{r}^{2}-\frac{A^{2}}{4\beta k^{2}}\right)\left(\text{Sinh}(\chi)\right)^{2} >0 \quad \forall \quad \chi>\hat{\chi},   
\end{equation}
and it is also bounded (\ref{eq69}) for all $\chi<\hat{\chi}$. Henceforth, we have a regular metric on the throat. On the other hand, away from the throat, the expression (\ref{eq72}) has a singularity when $\chi\rightarrow+\infty$, corresponding to $r\rightarrow+\infty$. It is an essential singularity of the metric (\ref{eq71}). This behavior is shown on Fig. \ref{fig3}, where it is clear that for points $\chi>\hat{\chi}$ the function $N$ tends to zero, while the function $f$ tends to infinite.

\begin{figure}[H]
\centering
\includegraphics[width=0.32\textwidth]{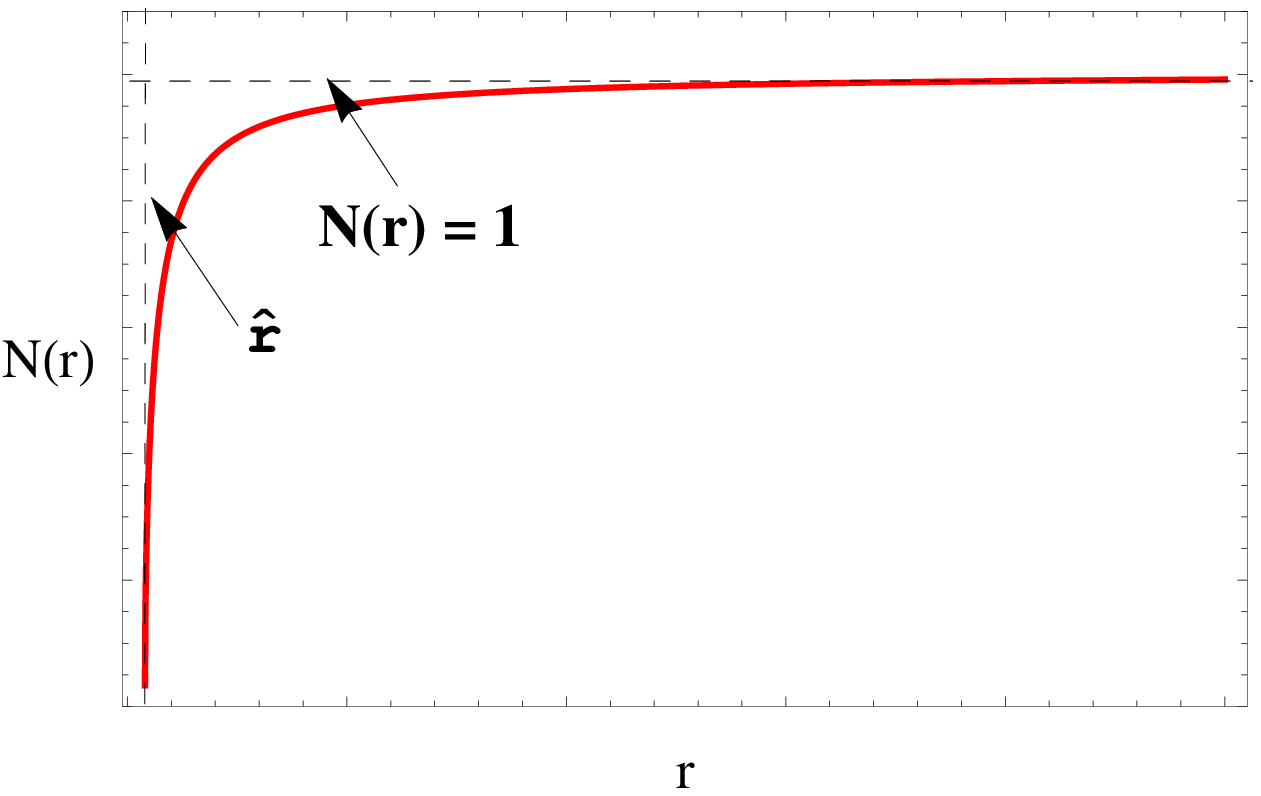}
\includegraphics[width=0.32\textwidth]{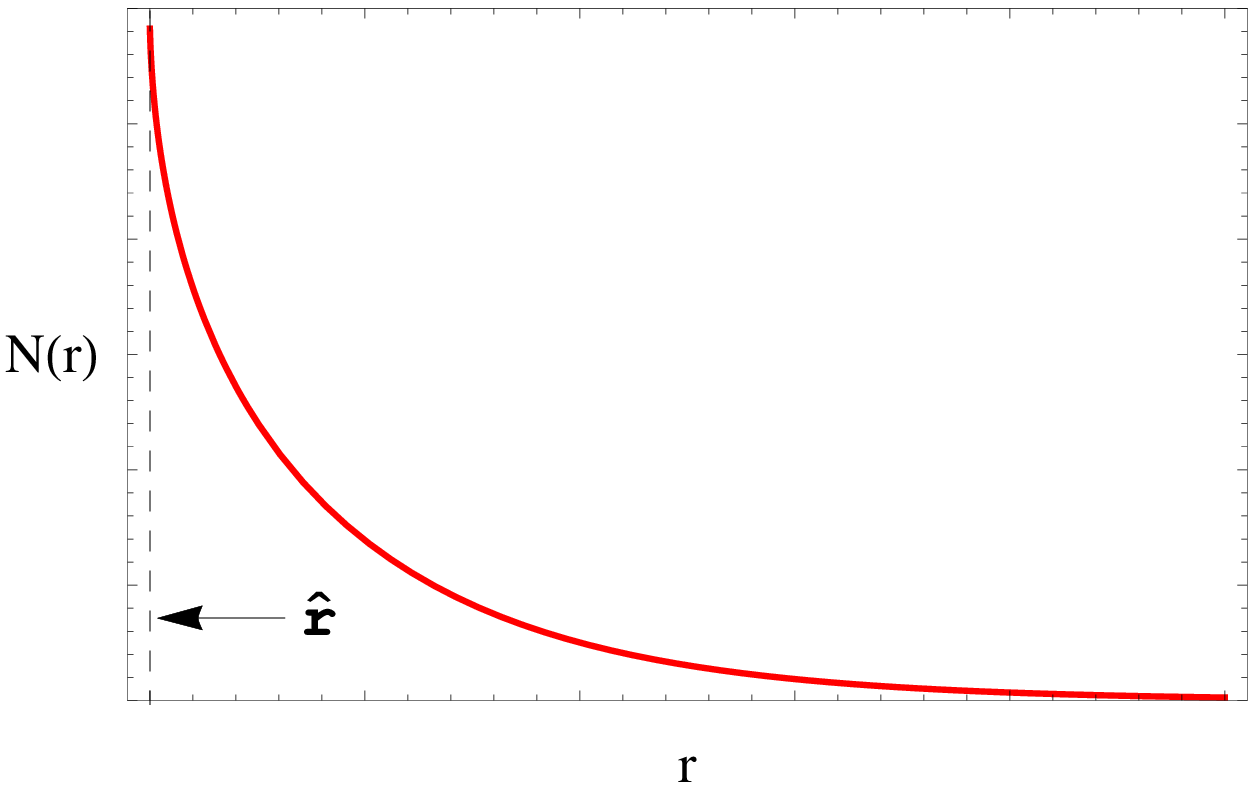}
\includegraphics[width=0.32\textwidth]{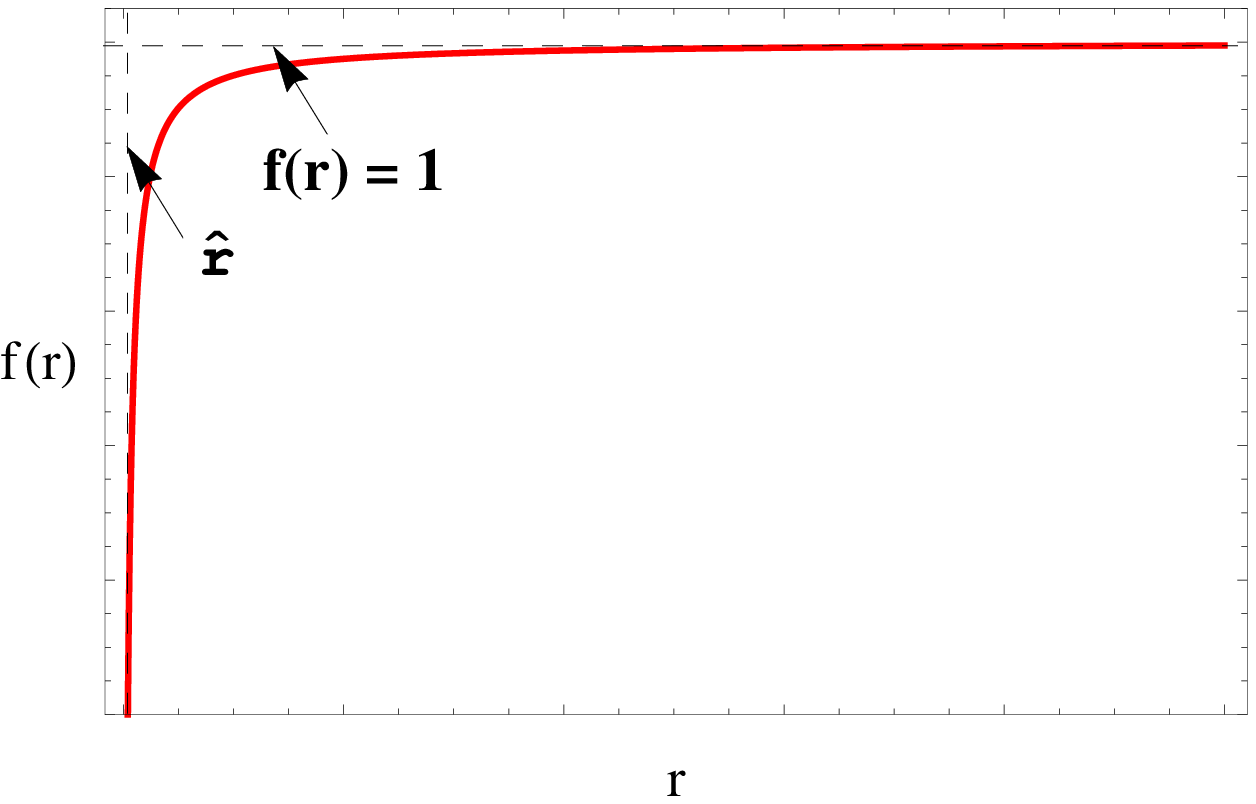}
\caption{
 \textbf{Left panel:} The lapse function $N$ against the radial coordinate $r$, for points $\chi<\hat{\chi}$ (dashed black line). As can be seen, this function shows an asymptotically flat behavior $N\rightarrow1$ when $r\rightarrow+\infty$. \textbf{Middle panel:}  The lapse function $N$ versus the radial coordinate $r$, for points $\chi>\hat{\chi}$ (dashed black line). As it is appreciated, $N$ tends to $0$ making the metric (\ref{eq71}) singular. \textbf{Right panel:} The function $f$ against the radial coordinate $r$, for points $\chi<\hat{\chi}$ (dashed black line). As can be seen, this function shows an asymptotically flat behavior $f\rightarrow1$ when $r\rightarrow+\infty$. 
}
\label{fig2}
\end{figure}

On the other hand, the functions $N$ and $f$ of  the metric (\ref{eq71}) in terms of the $\chi$ variable are depicted in Fig. \ref{fig3}, left and right panel respectively.

At this point it is illustrative to compare the charged throat solution with well--known 
Morris--Thorne (MT) WH space--time \cite{m1,m2,m3}. Of course, the comparison can be done only from the geometrical point of view.

Before going into the comparison, it is instructive to briefly go over the characteristics of the MT metric. So, in terms of the proper radial distance $l$, the most general spherically, symmetric, static and non--rotating WH space--time connecting two asymptotically flat regions is being describing by \cite{m1,m2,m3}
\begin{equation}\label{eq76}
ds^{2}_{\text{MT}}=-e^{-\phi(l)}dt^{2}+dl^{2}+r^{2}(l)\left[d\theta^{2}+\text{sin}^{2}\theta d\varphi^{2}\right],
\end{equation}
where the proper length $l$ and functions $\phi(l)$ and $r(l)$ should satisfy some requirements:
\begin{itemize}
    \item The coordinate $l$ covers the entire range  $(-\infty,+\infty)$.
    \item The assumed absence of event horizons implies that $\phi(l)$ must be everywhere finite. 
    \item In order for the spatial geometry to tend to an appropriate asymptotically flat limit one must impose, 
    \begin{equation}
        \lim_{l\rightarrow\pm\infty}\frac{r(l)}{|l|}=1 \quad \mbox{and} \quad \lim_{l\rightarrow\pm\infty}\phi(l)=\phi_{\pm},
    \end{equation}
    where $\phi_{\pm}$ must be finite.
    \item The radius of the WH throat is defined by 
    \begin{equation}
        \hat{r}=\text{min}\{r(l)\},
    \end{equation}
    where without of generality one can take the WH throat to occurs at $l=0$.
\end{itemize}
The explicit relation between the radial coordinate $r$ and the proper radial distance $l$ is given by
\begin{equation}\label{mtt}
    r(l)=\sqrt{\varrho^{2}+l^{2}},
\end{equation}
being $\varrho$ a constant. Then it is no hard to note that when $l\rightarrow-\infty$ then $r\rightarrow+\infty$ and when $l\rightarrow+\infty$ then $r\rightarrow+\infty$. In the present solution, the equivalence situation corresponds to take the $\chi$ coordinate as the proper radial distance $l$ in the MT space--time. However there is a small difference, the range that $\chi$ covers is $[0,+\infty)$. Although this is not an impediment to compare both models. As was pointed out earlier, when $\chi\rightarrow0^{+}$ the radial coordinate goes to $r\rightarrow+\infty$ and when $\chi\rightarrow+\infty$ the $r\rightarrow+\infty$, just as MT solution.   On the other hand, the WH throat of the MT solution is located when $l\rightarrow0$ yielding from Eq. (\ref{mtt}) to 
\begin{equation}
r(0)=|\varrho|,    
\end{equation}
$|\varrho|$ is identified as $r(0)=\hat{r}$ the WH throat size. Now, for the charged HL throat this occurs when $\chi\rightarrow\hat{\chi}$, hence $r(\hat{\chi})=\hat{r}$. As can be seen from the line element (\ref{eq76}) the solution is completely regular when it is described in term of the proper radial distance. Of course, replacing the value $l=0$ in (\ref{eq76}) and fixing both the time and the equatorial angle \i.e, $\theta=\pi/2$ the metric (\ref{eq76}) reduce to
\begin{equation}
    ds^{2}_{\text{MT}}=\hat{r}^{2}d\varphi^{2}.
\end{equation}
The above expression is describing the WH throat region, a circle with radius $\hat{r}$. Respect to the line element (\ref{eq71}) when the $\chi$ is used, the metric is completely regular. In fact, from Fig. \ref{fig4} it is clear that $g_{\chi\chi}$ metric component is completely regular and non vanishing at $\hat{\chi}$. On the other hand, if the MT solution is expressed in canonical or Schwarzschild--like coordinates, the line element (\ref{eq76}) becomes
\begin{equation}\label{eq81}
ds^{2}_{\text{MT}}=-e^{\Phi(r)_{\pm}}dt^{2}+\frac{dr^{2}}{1-\frac{b(r)_{\pm}}{r}}+r^{2}\left[d\theta^{2}+\text{sin}^{2}\theta d\varphi^{2}\right],    
\end{equation}
where the function $\phi(r)$ is identified as the red--shift function and $b(r)$ is the shape function. In this coordinates, the WH throat occurs at $b(\hat{r})=\hat{r}$ \cite{m1,m2,m3}. Therefore, it is clear from (\ref{eq81}) that $g_{rr}$ metric component is indeterminate making the metric degenerates. For our solution the same situation happens when the metric is expressed in terms of the radial coordinate $r$ \i.e., $\chi=\chi(r)$, where the line element (\ref{eq71}) degenerates. This fact is corroborated in right panel of Fig. \ref{fig3} where $f$ is vanishing at the throat $\hat{\chi}$. Hence the throat area is best described in terms of $l$ for the MT model and in terms of $\chi$ in the present case. In other words, $l$ and $\chi$ are the ``good coordinates'' when it comes to describing the throat in the mentioned models, respectively. Finally, in comparing the $g_{tt}$ components of the MT and (\ref{eq71}) solution, it is evident that the lapse function $N$ is playing the role of the ref--shift function $\phi$. Independently if one is on the chart described by $r$ or $\chi$, the lapse function $N$ is always regular and non vanishing at the throat, thus there is not event horizon as required (see Figs. \ref{fig2}--\ref{fig3}).

\begin{figure}[H]
\centering
\includegraphics[width=0.32\textwidth]{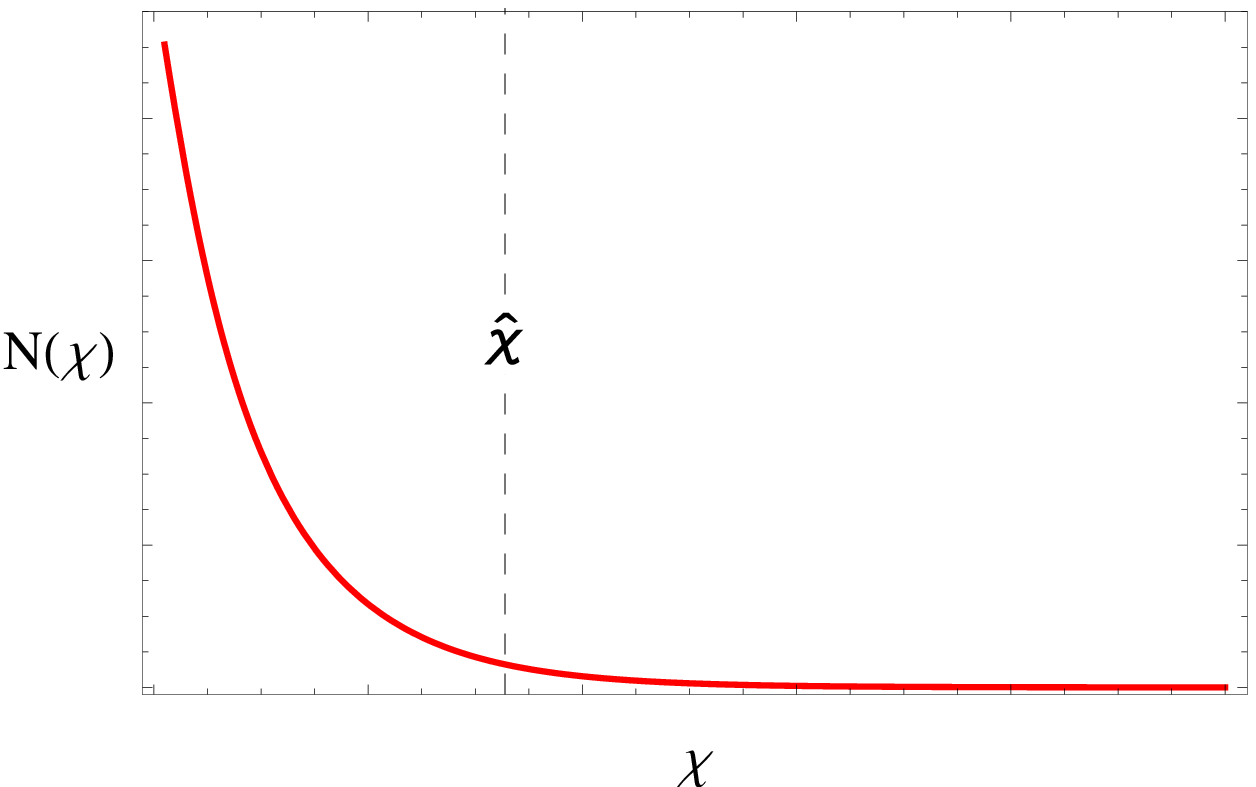}
\includegraphics[width=0.32\textwidth]{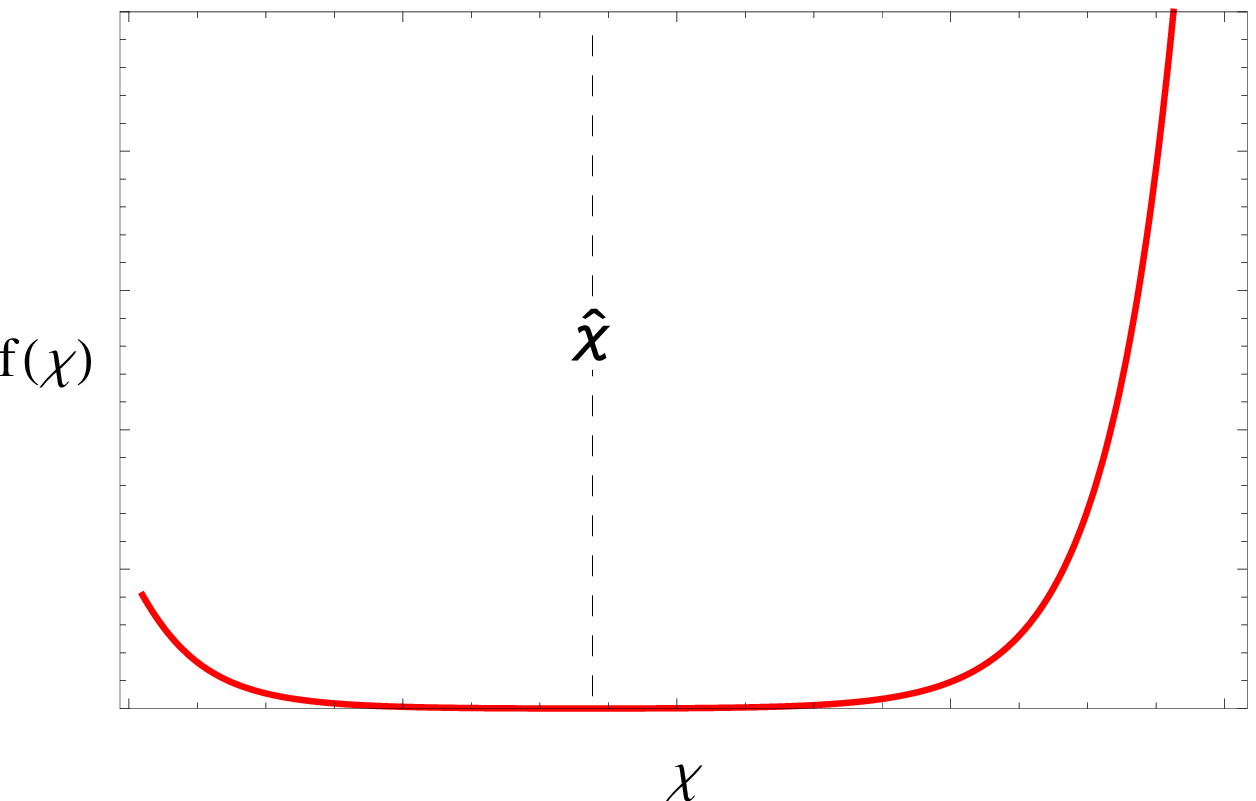}
\caption{
 \textbf{Left panel:} The lapse function $N$ trend versus the $\chi$ coordinate. As can be seen, this object is completely regular at the throat $\chi=\hat{\chi}$. \textbf{Right panel:} The behavior of the $f$ function against the $\chi$ variable. It is observed that at the throat, $f(\hat{\chi})=0$ making the metric (\ref{eq71}) singular.
}
\label{fig3}
\end{figure}

\begin{figure}[H]
\centering
\includegraphics[width=0.32\textwidth]{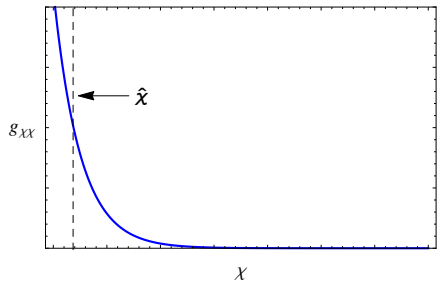}
\caption{
 The trend of the $g_{\chi\chi}$ metric component of the line element (\ref{eq71}) versus $\chi$.
}
\label{fig4}
\end{figure}

\section{Concluding remarks}\label{sec7}

We obtain a static spherically symmetric solution of the gauge vector--gravity HL equations \cite{r29,r30}, in the low energy regime where only the $z=1$ potential terms are considered. The solution depends on three parameters, the low energy couplings $\alpha$ and $\beta$ of the HL gravity and the electric charge. The metric describes a charged throat connecting a space--time which in the limit when $\alpha$ tends to 0 and $\beta$ to 1 becomes the Reissner--Nordstr\"{o}m solution of Einstein--Maxwell theory and an asymptotic singular manifold. The singularity is an essential one. 
To obtain the solution we introduce a new coordinate, which allows to describe it along the throat. We solve the field equations up to the nonlinear differential equation (\ref{eq35}). The solution is proven to exist starting from an initial data, satisfying the restriction (\ref{eq37}) at the throat and then showing its extension to an asymptotically flat space--time on one side and to an asymptotically singular manifold on the other side. We also describe the asymptotic behavior on the regular side and relate it to the initial data. We may also identify the coefficient $a$ of the asymptotic expansion with $M$ the mass of the system and the integration constant $A$ with $Q$ the electric charge.
From (\ref{eq37}) and (\ref{eq64}) we conclude that the solution exists when $M$ satisfies the restriction
$M>|A/ 2 k^{2} \beta^{1/2}|$.       
We notice that in the relativistic limit the restriction reduces to 
$M> |A/ 2|$
the condition in the Reissner--Nordstr\"{o}m solution of GR to have two horizons, the event horizon and the internal Cauchy horizon.
We have then obtained the anisotropic generalization, in the HL framework, of the Reissner--Nordstr\"{o}m solution of GR. Hence, although there are not horizons in the HL solution the physical relevant relation between the mass and the charge is also obtained as in the relativistic case. Although we have shown the existence of the solution and characterized it by using bounds and estimates, we do not have an explicit formula for the metric. We thus explicitly obtain the solution by following a numerical approach. The numerical solution of (\ref{eq35}) is depicted by the left panel of Fig. \ref{fig1}. The blue curve shows the existence of the throat located at the point $(\hat{\chi},\hat{r})$, whilst the red curve displays the GR limit. On the other hand the middle and right panels in Fig. \ref{fig1} exhibit the behavior of $\chi$ is terms of $r$. The one to one correspondence is corroborated and it is clear from these panels, that the $\chi$ variable is the good one to describe the throat region, since $r(\chi)$ has a global minimum at the throat. As was mentioned in the relativistic limit, the asymptotic behavior of this solution yields the well--known Reissner--Nordst\"{o}m solution. The asymptotic solution is given by Eqs. (\ref{eq56}) and (\ref{eq59}), where in the relativistic limit \i.e, $\beta\rightarrow 1$ and $\alpha\rightarrow 0$, the parameters $a$ and $b$ can be interpreted as $-2M$ and $Q^{2}$ respectively, where $M$ is Schwarzschild mass and $Q$ the electric charge. 

In considering the full metric given by Eq. (\ref{eq71}), its graphical representation in Fig. \ref{fig2} depicts the behavior for $N$ and $f$ as functions of the radial coordinate $r$. Interestingly, both functions for points $\chi<\hat{\chi}$ tend to one, hence the metric (\ref{eq71}) is asymptotically flat, describing a Minkowski space--time. Those are the reasons why we called the solution a throat and not a WH, because following the definition given in \cite{m1,m2,m3} a WH structure is asymptotically flat for points below and above the throat. In this case we have an essential singularity on one side and an asymptotically flat region on the other side.
On the other hand, the line element expressed in terms of $\chi$ allows to compare with the usual WH behavior at the throat. In fact, the Fig. \ref{fig3} depicts the trend of $N$ and $f$ as functions of $\chi$, as can be observed $N$ at $\hat{\chi}$ is not zero while $f$ is. The geometrical behavior is exactly the same as the MT solution \cite{m1,m2,m3} where a traversable WH solution must satisfy $\phi(\hat{r})\neq0$ and $b(\hat{r})=\hat{r}$. Of course, it is evident from (\ref{eq81}) that evaluating the shape function $b(r)$ at the WH throat, the line element becomes singular. 

Finally, it is remarkable to point out that this kind of solution are characteristic of the non--projectable HL theory, without introducing any kind of exotic matter distribution \cite{r31} as it is required in the framework of GR \cite{m1,m2,m3}. The throat is supported by the static electric field, introduced by the 3+1 dimensional reduced theory \cite{r29,r30} after performing the Kaluza--Klein program, acting as matter content. Concerning this point, the energy conditions, Bianchi's identities for the matter sector and other treatment performed in relativistic theories when considering this type of solutions, are still not known in the framework of the HL theory.

\section*{Acknowledgements}
F. Tello-Ortiz thanks the financial support by the CONICYT PFCHA/DOCTORADO-NACIONAL/2019-21190856 and project ANT-1956 at the Universidad de Antofagasta, Chile.

\end{document}